\documentclass[aps,pre,twocolumn,showpacs,amsmath,amsfonts,preprintnumbers]{revtex4}
\usepackage{epsfig}
\usepackage{yfonts}
\usepackage{amssymb}
\usepackage{graphicx}
\swabfamily
\newcommand{\Op}[1]{{{\mathrm{\hat{#1}}}}}

\begin{document}
\title{Noise and Controllability: suppression of controllability in large quantum  systems}
\author{M. Khasin$^{1}$ and R. Kosloff$^{2}$ }

\affiliation{$^{1}$Department of Physics and Astronomy, Michigan State University, East Lansing, MI 48824 USA}

\affiliation{$^{2}$Fritz Haber Research Center for Molecular Dynamics, 
Hebrew University of Jerusalem, Jerusalem 91904, Israel}

\date{\today }

\begin{abstract}

A closed quantum system is defined as completely controllable if an arbitrary unitary transformation can be executed using the available controls.
In practice,  control fields are a source of unavoidable noise. 
Can one design control fields  such that the effect of noise is negligible on the time-scale of  the transformation?
Complete controllability in practice requires that the effect of noise can be suppressed for an arbitrary transformation.
The present study considers a paradigm of control, where the Lie-algebraic structure of the control Hamiltonian is fixed, while the size of the system increases, determined by the dimension of the Hilbert space representation of the algebra. 
We show that for large quantum systems, generic noise in the controls  dominates  for a  typical class of target transformations i.e., complete controllability is destroyed by the noise.
\end{abstract}
\maketitle

Coherent control was constructed to steer a quantum system from an initial state to a target state via an external field \cite{rice92,shapiro03}.  The idea is to control the interference pathway governing the dynamics.
For pure initial and final (target) states the method can be termed \textit{state-to-state} coherent control. A generalization is steering simultaneously a set of initial pure states to a set of final states, i.e. controlling  
a unitary transformation. Such an application sets the foundation for a quantum gate operation \cite{tesch01,k181,rabitz05}.
Three basic questions  address feasibility of coherent control. The first, for a preset initial  and target state does a control filed exist?
This is the problem of controllability. The second, how to construct a field that  leads to the target? This is the problem of synthesis.
The third, how to optimize the field that carries out this task?
This is the problem of Optimal Control Theory \cite{rabitz88,k67,k193}. 
Experimentally there has been a remarkable success in constructing devices designed to generate arbitrary control fields \cite{tom99,brixner07,gerber08}.  
Nevertheless, in practice controllability is hard to achieve even for small quantum systems \cite{glaser09,glaser07,herr07}.
Applications toward quantum information processing require  upscaling of the control procedures 
to  large quantum  systems.
The issue of  controllability of a closed quantum system has been addressed by Tarn and Clark \cite{tarn}. Their theorem states that for a finite dimensional closed quantum system, if the  control operators and the unperturbed Hamiltonian  generate the Lie-algebra of all Hermitian operators,  the system is \textit{completely controllable}, i.e., an arbitrary unitary transformation of the system can be realized by an appropriate application of the controls \cite{rama}.  
Complete controllability implies {state-to-state controllability}.

In practice the controlled systems are open and the system-bath coupling introduces noise into the system dynamics. 
A number of techniques have been designed to combat the  environmental noise and effective and ingenious  algorithms have been invented and explored \cite{viola99,viola09,lidar05,lidar07,gershon07}. 
The present study focuses on the effect of the noise originating in the control field. The magnitude of this noise
depends on the properties of the control field. This dependence raises a fundamental question: is it always possible, for a given target, to design a field, such that the effect of the associated noise can be neglected? This problem has been extensively investigated in the context of fault-tolerant quantum computation \cite{preskill98}, where various schemes have been designed to fight the noise in the gates. 
In quantum computation the number of gates  increases  with the size of the system. In many fields, for example in  NMR  \cite{nielsen10} or in control of molecular systems \cite{rice92} a different control paradigm is standard. There the control operators are fixed while the size of the system may vary.
This is the control paradigm considered in the present study.
The presentation is restricted to quantum system with finite Hilbert space dimension. It is assumed that the control operators are elements of the  spectrum-generating algebra of the quantum system \cite{bohm}. For a finite dimensional system it is sufficient to consider a compact semisimple algebra \cite{gilmorebook}. The size of the system is determined by the dimension of the Hilbert-space representation of the algebra. It can be characterized by a parameter $N$, which is a scaling factor of the highest weight of the representation  \cite{Gilmore79}. The physical interpretation of $N$ depends on the system. It can be, for example, a number of particles in the system or the number of energy levels.
For this  paradigm it is shown that complete controllability of a large quantum system, i.e., for $N \gg 1$,
even in the weaker sense of state to state control, does not survive in the presence of generic noise in the control fields. As a result the control is not scalable with the size of the system.

 Let the Hamiltonian of the controlled system be
\begin{equation}
\Op H=\Op H_0+\sum_k \left[u_k(t)+ \xi_k(t)\right]\Op X_k.
\label{eq:1}
\end{equation}
where $u_k(t)$ are control fields and the noise  $\xi_k(t)$  is delta-correlated Gaussian noise  with 
$\left\langle \xi_k(t)\xi_l(t')\right\rangle = 2\eta |u_k(t)| \delta_{kl} \delta(t-t')$. The dimensionless number $\eta\ge0$ measures the relative strength of noise in the control fields.
The operators $\Op X_k$ are elements of the spectrum generating algebra associated with $\Op H_0$ and are termed controls in what follows.
The equation of motion for the density operator of the system is given by \cite{gorini76}
\begin{eqnarray}
\frac{\partial}{\partial t}\Op \rho &=&-i \left[\Op H_0+\sum_k u_k(t)\Op X_k, \Op\rho \right] \nonumber \\
&-&\eta \sum_k  |u_k(t)|\left[\Op X_k,\left[\Op X_k,\Op\rho \right]\right], \label{liouville}
\end{eqnarray}

In the absence of noise the system (\ref{liouville}) is completely controllable for a generic $\Op H_0$.
Due to the noise the purity ${\cal P} \equiv \texttt{Tr}\left\{ \Op \rho^2\right\}$ of an initially pure state $\Op \rho=\left|\psi\right\rangle\left\langle \psi\right|$   will decrease. The effect of noise can be neglected only if the purity loss during the target transformation is small, i.e., $\Delta {\cal P} \ll 1$.   Otherwise, the
state-to-state and, hence, complete controllability is lost.
The instantaneous rate of purity loss for a pure state $\Op \rho=\left|\psi\right\rangle\left\langle \psi\right|$ evolving according to Eq.(\ref{liouville}) is given \cite{viola07} by
\begin{equation}
\Gamma\left[\psi\right] \equiv -\frac{d}{dt} \texttt{Tr}\left\{ \Op \rho^2\right\}|_{\Op \rho=\left|\psi\right\rangle\left\langle \psi\right|}=4 \eta\sum_{k, u_k \neq 0} |u_k(t)|\Delta_{\Op X_k}\left[\psi\right], \label{rateofloss}
\end{equation}
where $\Delta_{\Op X_k}\left[\psi\right]$ is the variance of the control operator $\Op X_k$ in the state $\psi$:
\begin{equation}
\Delta_{\Op X_k}\left[\psi\right]\equiv \left\langle \psi\left|\Op X_k^2\right|\psi \right\rangle-\left\langle \psi\left|\Op X_k\right|\psi \label{uncertainty} \right\rangle^2.
\end{equation}
It should be noted that the rate of purity loss   increases with the variance of $\Op X_k$.  A generic state of the system is characterized by $\Delta_{\Op X_k}\left[\psi\right] \sim N^2$.

The logic of the  presentation is as follows. First a class of state-to-state target transformations is defined. The transformations can be accomplished in the absence of noise due to complete controllability. The time duration of these transformations can be bounded from below. Next the lower bound of the purity loss rate for the evolving state in the presence of noise is estimated. For small purity loss  the two bounds can be combined to obtain the lower bound on the purity loss during the transformation. This bound depends on the relative strength of noise $\eta$. 
The crux of the class of transformations considered is that for \textit{any realization} of the control fields accomplishing the transformation the evolving system resides for a long period of time in states with large variance $\sim N^2$ with respect to the control operators. 
The large variance will generate  $\sim \eta N^2$ rate of purity loss in the presence of noise in the control fields. 
For this class of transformations it is found that the lower bound on the time of transformation scales as $\sim N^{-1}$, therefore, $\eta$ must be $O(N^{-1})$ in order that the purity loss be negligible. Since in practice the relative magnitude of the noise cannot be made arbitrarily small, it follows that the loss of purity cannot be neglected for large systems. As a result, a transformation from an arbitrary pure state to another  pure state cannot be accomplished.  
Large systems are not state-to-state controllable and are  therefore not completely controllable.

Let $\left|n\right\rangle$ denote the eigenstates of $\Op H_0$.
We consider the transformation from the initial state $\left|\psi_{i}\right\rangle=\sum_n r_{i,n} e^{i \phi_{i,n}}\left|n\right\rangle$ to the final state  $\left|\psi_f\right\rangle=\sum_n r_{f,n} e^{i \phi_{f,n}}\left|n\right\rangle$,  such that
\begin{equation}
1 \gg \left\|\Delta \textbf{r}\right\| \ge \varepsilon >0,\label{minchange}
\end{equation}
where  $\left\|\Delta \textbf{r}\right\|$ is the Euclidean norm of  $\Delta \textbf{r}\equiv \textbf{r}_{f}-\textbf{r}_i$, $\textbf{r}_{i}=(r_{i,1},r_{i,2},...)$ and $\textbf{r}_{f}=(r_{f,1},r_{f,2},...)$. 
The choice of norm excludes changes to states that can be reached by free propagation generated by  $\Op H_0$. As a consequence, the free evolution does not contribute to the bound on the time of the transformation $\psi_i\rightarrow \psi_f$. Under assumption that the noise is small the bound can be estimated in the zero order in the noise strength.
For estimation of the bound an auxiliary operator $\Op A$ is defined such that: (i) it commutes with $\Op H_0$; (ii) its expectation value changes during the transformation.  Since $\Op A$ commutes 
with $\Op H_0$  the change of its expectation value during the transformation is due to the operation of the control fields.
We define $\Op A=\sum_n s_n \left|n\right\rangle\left\langle n\right|$,
where $s_n=\texttt{sign} \left\{\Delta r_n\right\}$. The change of the expectation value of the operator $\Op A$ during the transformation $\psi_{i} \rightarrow \psi_{f}$ is given by
\begin{eqnarray}
\left\langle\Op A\right\rangle_f-\left\langle\Op A\right\rangle_i&=&\sum_n \left|\Delta r_n \right|\left(r_{i,n} +r_{f,n}\right)\nonumber \\
&\ge& \sum_n \Delta r_n^2=\left\|\Delta \textbf{r}\right\|^2, \label{sum2}
\end{eqnarray}
where we have used the fact that the vector of amplitudes $\textbf{r}$ is nonnegative, and, therefore, $\left|\Delta r_n\right|>r_{i,n}$ only if $\Delta r_n\ge 0$. 
Using inequality (\ref{minchange}) we obtain
\begin{equation}
\left\langle\Op A\right\rangle_f-\left\langle\Op A\right\rangle_i \ge \varepsilon^2, \label{minchange2}
\end{equation}
which gives the minimal change of the expectation value of the operator $\Op A$ during the transformation $\psi_{i}\rightarrow \psi_f$.
On the other hand, the change of the expectation value of $\Op A$ can be estimated from the Heisenberg equations:
\begin{eqnarray}
\frac{d}{dt}\Op A&=&i\sum_k u_k(t) \left[\Op X_k, \Op A\right]
\end{eqnarray}
where we have used the fact that $\left[\Op H_0,\Op A\right]=0$. Let the time of the transformation be $T$. Then,
\begin{eqnarray}
\left\langle\Op A\right\rangle_f&-&\left\langle\Op A\right\rangle_i=\int_0^{T}\frac{d}{dt}\left\langle \Op A \right\rangle \ dt \nonumber \\
&\le& \sum_k \int_0^{T}\left|u_k(t)\right| \ dt \max_{0\le t\le T}\left|\left\langle \left[\Op X_k, \Op A\right]\right\rangle \right|\  \nonumber \\
&\le& 2  \sum_k \int_0^{T}\left|u_k(t)\right| \ dt \left|\Lambda_{k}\right|, \label{minchange3}
\end{eqnarray}
where $\Lambda_k \sim N$ stands for the  eigenvalue of $\Op X_k$, maximal by the absolute value.  In the derivation we have used the fact that the eigenvalues of $\Op A$ are $\pm 1$.   Defining the average control amplitude $\bar{u}_k\equiv \frac{1}{T}\int_0^{T}\left|u_k(t)\right| \ dt$,
 and using Eqs.(\ref{minchange2}) and (\ref{minchange3}), we arrive at the inequality
\begin{eqnarray}
T\ge& \varepsilon^2 \left(2\sum_k \bar{u}_k \left|\Lambda_{k}\right|\right)^{-1} \sim \varepsilon^2 \left(2N\sum_k \bar{u}_k\right)^{-1} , \label{timebound2}
\end{eqnarray}
which bounds the time of the transformation for given $\bar{u}_k$. This bound is similar to bounds obtained for the transformation to an orthogonal state in Refs.\cite{Margolus98}.

Without loss of generality we assume that the time $T$  of the transformation $\psi_{i}\rightarrow \psi_f$ is the first passage time when the evolving state $\left|\psi(t)\right\rangle=\sum_n r_n(t) e^{i {\phi}_n(t)}\left|n\right\rangle$ "crosses the border" $\left\|\textbf{r}(T)-\textbf{r}_i\right\|= \varepsilon$, i.e., such that for $t<T$ we have $\left\|\textbf{r}(t)-\textbf{r}_i\right\|<\varepsilon$.
Under assumption that the purity loss $\Delta \cal{P}$ during the transformation is small, the evolving state can be approximated by $\rho(t)=\rho^{(0)}+\rho^{(1)}\approx \rho^{(0)}=\left|\psi(t)\right\rangle\left\langle \psi(t)\right|$.  Taking the leading contribution of $\rho^{(1)}$ into account, we estimate the lower bound on the purity loss from Eq.(\ref{rateofloss}):
\begin{eqnarray}
\Delta {\cal{P}}&\ge& 4 T\eta \sum_k   \bar{u}_k \min_{0\le t \le T}\{\Delta_{\Op X_k}\left[\psi(t)\right]\nonumber \\
&+&\frac{1}{2}\left\langle \psi(t)\left|[\Op X_k,[\Op X_k,\rho^{(1)}(t)]]\right|\psi(t)\right\rangle\}. \label{minpur}
\end{eqnarray}
We further assume that during the transformation $\Delta_{\Op X_k}\left[\psi(t)\right] \sim (\Lambda_k)^2 \sim N^2$.
In this case we can neglect the $\rho^{(1)}$-dependent term in the inequality (\ref{minpur}). Using the inequality (\ref{timebound2}), we obtain
\begin{equation}
\Delta {\cal{P}}\ge \frac{2\varepsilon^2 \eta \sum_k   \bar{u}_k\min_{0\le t \le T}\left\{\Delta_{\Op X_k}\left[\psi(t)\right]\right\}}{\sum_k \bar{u}_k \left|\Lambda_{k}\right|}. \label{minpur2}
\end{equation}

To estimate $\min_{0\le t \le T}\left\{\Delta_{\Op X_k}\left[\psi(t)\right]\right\}$ we find the lower bound on the  variance of $\Op X_k$ in the states $\left|\psi\right\rangle=\sum_n r_n e^{i {\phi}_n}\left|n\right\rangle$  such that $\left\|\textbf{r}-\textbf{r}_i\right\|\le \varepsilon$.
The variance $\Delta_{\Op X_k}\left[\psi\right]$ is a function of the amplitudes $\textbf{r}=(r_1,r_2,...)$ and the phases $\phi_1,\phi_2,...$. 
The free evolution can change the phases at no cost in purity. Therefore, the minimal variance attainable for given amplitudes is sought:
\begin{equation}
\tilde{\Delta}_{\Op X_k}\left(\textbf{r} \right)\equiv \min_{\phi_1,\phi_2,...}\left\{\Delta_{\Op X_k}\left[\psi\right]\right\} \label{min}
\end{equation}
We assume, that $\tilde{\Delta}_{\Op X_k}\left(\textbf{r} \right)$ is a smooth function of $\textbf{r}$ for $\left\|\textbf{r}-\textbf{r}_i\right\|\le \left\|\Delta \textbf{r}\right\|$, i.e., for sufficiently small $\Delta \textbf{r}$ and $|\delta \textbf{r}|\le \left\|\Delta \textbf{r}\right\|$ we can expand $\tilde{\Delta}_{\Op X_k}\left(\textbf{r}_i+\delta \textbf{r} \right)\approx \tilde{\Delta}_{\Op X_k}\left(\textbf{r}_i\right)+\nabla \tilde{\Delta}_{\Op X_k}\left(\textbf{r}_i\right)\cdot \delta \textbf{r}$.
We note that
\begin{equation}
\left|\delta \tilde{\Delta}_{\Op X_k}\left(\textbf{r}\right)\right|\equiv\left|\nabla \tilde{\Delta}_{\Op X_k}\left(\textbf{r}\right)\cdot \delta \textbf{r}\right|\le \left\|\nabla \tilde{\Delta}_{\Op X_k}\left(\textbf{r}\right)\right\|\left\|\delta \textbf{r}\right\| \label{change}.
\end{equation}
Let the minimum in Eq.(\ref{min}) obtain at $\phi_1^*(\textbf{r}),\phi_2^*(\textbf{r}),...$. Let us denote the associated state as $\psi^*$. Then 
$\tilde{\Delta}_{\Op X_k}\left(\textbf{r} \right)=\Delta_{\Op X_k}\left[\psi^*\right]$
and
\begin{eqnarray}
\nabla \tilde{\Delta}_{\Op X_k}\left(\textbf{r} \right)&=&\nabla \Delta_{\Op X_k}\left[\psi^*\right] \label{nabla}
\end{eqnarray}
It should be noted that $\Delta_{\Op X_k}\left[\psi^*\right]$ depends on $r_n$ both through the amplitudes of $\psi^*$ and through the phases $\phi_n^*(\textbf{r})$, which are also functions of $\textbf{r}$. Nonetheless, since $\phi_n^*(\textbf{r})$ are defined as giving the minimum of $\Delta_{\Op X_k}\left[\psi\right]$, derivatives of $\Delta_{\Op X_k}\left[\psi^*\right]$ with respect to $\phi_n^*(\textbf{r})$ vanish and $\phi_n^*(\textbf{r})$ may be considered as $\textbf{r}$-independent for the operator $\nabla$ in the rhs of Eq.(\ref{nabla}). 

Using the definition Eq.(\ref{uncertainty}) we obtain 
\begin{eqnarray}
\nabla \tilde{\Delta}_{\Op X_k}\left(\textbf{r} \right)&=&\nabla \left\langle \psi^*\left|\Op X_k^2\right|\psi^* \right\rangle\nonumber \\
&-&2 \left\langle \psi^*\left|\Op X_k\right|\psi^* \right\rangle \nabla \left\langle \psi^*\left|\Op X_k\right|\psi^* \right\rangle.\label{inter}
\end{eqnarray}
Using the explicit form of $\psi^*$ and the fact that $\nabla$ act only on the state's amplitudes we  can show that the Euclidean norm of the rhs of Eq.(\ref{inter}) is bounded by $3\sqrt{2}\left(\Lambda_k\right)^2$. Then,  Eqs.  (\ref{change}) and (\ref{inter}) imply
$\left|\delta \tilde{\Delta}_{\Op X_k}\left(\textbf{r}\right)\right|\le 3\sqrt{2}\left(\Lambda_k\right)^2\left\|\delta \textbf{r}\right\|$.
It follows that for  $\left|\psi\right\rangle=\sum_n r_n e^{i {\phi}_n}\left|n\right\rangle$ with $\left\|\textbf{r}-\textbf{r}_i\right\|\le \varepsilon \ll 1$ 
\begin{equation}
{\Delta}_{\Op X_k}\left[\psi\right]\ge \tilde{\Delta}_{\Op X_k}\left(\textbf{r}_i\right)-3\sqrt{2}\left(\Lambda_k\right)^2\varepsilon \label{minun}
\end{equation}
From inequalities (\ref{minpur2}) and (\ref{minun}) we obtain
\begin{eqnarray}
\Delta {\cal{P}}&\ge& 2 \varepsilon^2   \eta \left(\frac{ \min_l\left\{\tilde{\Delta}_{\Op X_l}\left(\textbf{r}_i\right)-3\sqrt{2}\left(\Lambda_{l}\right)^2\varepsilon \right\}}{ \max_l\left\{\left|\Lambda_{l}\right|\right\}}\right) \label{gen0}
\end{eqnarray} 
The variance $\Delta_{\Op X_k}\left[\psi\right]$ scales as $N^2$ in a generic state of the system . The scaling of $\tilde{\Delta}_{\Op X_k}\left(\textbf{r}_i\right)$ is a more subtle question, since it is the outcome of the minimization with respect to the phases in the eigenstates basis. For our purposes it is sufficient to show that $\tilde{\Delta}_{\Op X_k}\left(\textbf{r}_i\right)\sim N^2$ for \textit{some} $\left|\psi_i\right\rangle$. Let's consider a generic eigenstate $\left|\phi\right\rangle$ of $\Op H_0$. The variance $\Delta_{\Op X_k}\left[\phi\right]$ scales as $N^2$ for all $k$. Moreover, the variance is independent of phases. Therefore, taking $\left|\psi_i\right\rangle=\left|\phi\right\rangle$ we shall have $\tilde{\Delta}_{\Op X_k}\left(\textbf{r}_i\right)\sim N^2$, and, for sufficiently small $\varepsilon$, inequality (\ref{gen0}) will imply
\begin{eqnarray}
\Delta {\cal{P}}&\ge& 2 \varepsilon^2   \eta N \label{gen}.
\end{eqnarray}
It is important to note that inequality (\ref{gen}) holds for small $\varepsilon$, which can still be of the order of unity with respect to $N$. As a consequence, for such target transformations the rhs of inequality (\ref{gen}) will scale as $\eta N$. In order that the purity loss be negligible, the relative noise strength $\eta$ must scale as $O(N^{-1})$, which is unrealistic for  large $N$.

As an example, let's consider a system of cold  atoms in a double well trap \cite{anglin01}. The Hamiltonian of the system can be put in the form
\begin{eqnarray}
\Op H_0=-\omega \Op J_x+\delta \Op J_z+ \frac{U}{N} \Op J_z^2 \label{ham},
\end{eqnarray}  
where $\Op J_k$ are elements of the $su(2)$ Lie-algebra. The operator $\Op J_z$ corresponds to the population difference between the wells and $\Op J_y$ to the population flow between the wells. The on-site interaction strength is given by $U$,  $\omega$ is the hopping rate of the atoms and $\delta$ determines the tilt of the potential wells. $N$ is the number of atoms, corresponding to the $N+1$ dimensional  irreducible representation of the $su(2)$ algebra, with the spin number $j=N/2$. Control can be attained  via the control operators  $\Op J_x$ and $\Op J_z$, i.e., by modulating the hopping rate and the tilt of the wells \cite{Steinhauer}.  For a typical  system of $N\gtrsim 10^5$ atoms the necessary condition $\eta N \sim 1$ cannot be satisfied in practice. Therefore, the purity loss in the system during a target transformation of the type considered in the present work will be of the order of unity and the effect of noise cannot be neglected.

We conclude that the state-to-state controllability of large quantum systems  is destroyed by the noise on the control. The strategy of suppressing the influence of the noise by a faster control will fail due to the required increase in field amplitude, inevitably accompanied by the increase in noise. The purity loss of the evolving state will be faster due to the increase in noise on the controls.
The order-of-unity decrease of the purity  implies that the relative error in the expectation values of some of the system operators in the target state is of the order of unity. This large relative error is characteristic for transformations between states, where variance of the control operators is $\sim N^2$. It is expected that the relative error is small for target transformations between the "classical" generalized coherent states \cite{Zhang} with respect to the spectrum generating algebra, where the maximal variance of the control operators is $\sim N$.  An example is transformations between the spin-coherent states \cite{arecchi} of a quantum spin, corresponding to  BEC states of atoms in the double-well trap. 
  Finally, it is interesting to note that the scalability of the control paradigm considered in the present study can be attacked from a different angle of resource management in quantum computing \cite{blume}. Remarkably, the notion of what constitutes a large quantum system is similar in both studies.  \paragraph*{Acknowledgements}
  We are grateful to Kalvi institute University of Santa Barbara for hospitality. Work supported by the Israeli Science foundation.

%\bibliography{C:/Users/user/Desktop/Documents/papers/controlability/control,C:/Users/user/Desktop/Documents/papers/pub,C:/Users/user/Desktop/Documents/papers/pumpprob} 
%\bibliography{control,/Users/ronnie/Desktop/LINUX/Text/Database/pub}
\end{document}